\documentclass[a4paper,11pt]{article}
\usepackage{pos}
\usepackage{subfigure}
\usepackage{overpic}
\usepackage{enumitem}
\title{Simulations of cross media showers with CORSIKA 8}
 \ShortTitle{Cross media showers with C8}

\author*[a]{Juan Ammerman-Yebra}
\author[b]{Uzair Abdul Latif}
\author[c]{Nikolaos Karastathis}
\author{Tim Huege$^{c,d}$ for the CORSIKA 8 Collaboration}
\author[b]{Simon de Kockere}

\affiliation[a]{Instituto Galego de F\'isica de Altas Enerx\'ias, Universidade de Santiago de Compostela,\\
  R\'ua Xoaqu\'in D\'iaz de R\'abago, 15705, Spain}
\affiliation[b]{Vrije Universiteit Brussel, Dienst ELEM, Inter-University Institute for High Energies (IIHE),\\
Pleinlaan 2, 1050 Brussels, Belgium}

\affiliation[c]{Institute for Astroparticle Physics (IAP), Karlsruhe Institute of Technology,\\
P.O. Box 3640, 76021 Karlsruhe, Germany}
\affiliation[d]{Vrije Universiteit Brussel, Astrophysical Institute,\\
Pleinlaan 2, 1050 Brussels, Belgium}

\emailAdd{juan.ammerman.yebra@usc.es}
\emailAdd{Uzair.Abdul.Latif@vub.be}
\emailAdd{nikolaos.karastathis@kit.edu}
\emailAdd{tim.huege@kit.edu}
\emailAdd{Simon.De.Kockere@vub.be}

\abstract{The CORSIKA 8 project aims to develop a versatile and modern framework for particle shower simulations that meets the new needs of experiments and addresses the caveats of existing codes. Of particular relevance is the ability to compute particle showers that pass through two or more different media, of varying density, in a single run within a single code. CORSIKA 8 achieves this flexibility by using a volume tree that specifies volume containment, allowing one to quickly query to which medium a point belongs. Thanks to this design we are able to construct very specific environments with different geometries and media. As an example, we demonstrate this new functionality by running particle showers penetrating from air into Antarctic ice and validating them with a combination of the well-established CORSIKA 7 and \textsc{Geant4} codes.}

\ConferenceLogo{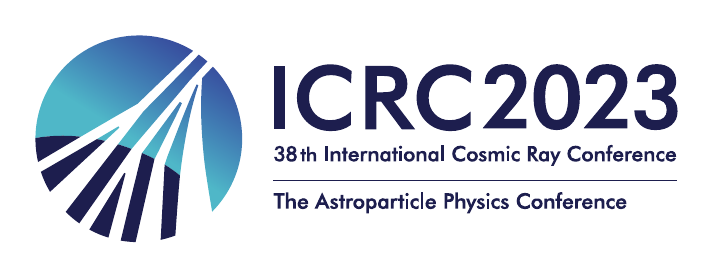}

\FullConference{%
38th International Cosmic Ray Conference (ICRC2023)\\
  26 July - 3 August, 2023\\
  Nagoya, Japan}

\begin{document}
\maketitle

\section{Introduction}
The detection of ultra-high energy cosmic rays and the development of new experiments has always gone hand in hand with particle shower simulations. Until now the two main air shower Monte Carlo, CORSIKA 7\cite{CORSIKA} and AIRES \cite{Sciutto:1999jh}, have provided a wonderful output but, being designed and optimized for air showers, lack some of the features required for modern experiments. Currently, IceCube \cite{halzen2005icecube} and KM3NeT \cite{distefano2009km3net} already need Monte Carlos that handle cross-media showers for background rejection and new proposals exploring tau-induced showers, like TAMBO \cite{romero2020andean}, would benefit from freely configurable media.

The CORSIKA 8 project aims to develop a new Monte Carlo for particle showers designed to tackle some of the current issues in the existing Monte Carlos. The project is a community-driven effort to establish a common framework for the simulation of particle showers. Being developed in modern C++ it aims to achieve an equilibrium between high-performance, modularity and easy maintainability.

Previous air shower Monte Carlo designed in the 90's did not have in mind the current computational options and new experimental needs. Some of these limitations were:
\begin{itemize}
    \item Not designed to take into account new computational advancements like GPU parallelization.
    \item Limited to the propagation of particles in one medium in one run, except for \textsc{Geant4}, which can handle different homogeneous media.
    \item Difficult integration of new features or interaction models.
    \item New modules cannot modify the particle propagation, of interest for radio or Cerenkov light emission.
\end{itemize}

The current version of CORSIKA 8 already tackles most of these problems \cite{huege_icrc2023}. This proceeding will focus on the capabilities of CORSIKA 8 to simulate cross-media showers and the new options this feature offers. 

\section{Cross-media showers}
\subsection{The CORSIKA 8 world structure}
\begin{figure}[ht]
\centering
\includegraphics[width=1.0\linewidth,trim={0 2cm 0 2cm},clip]{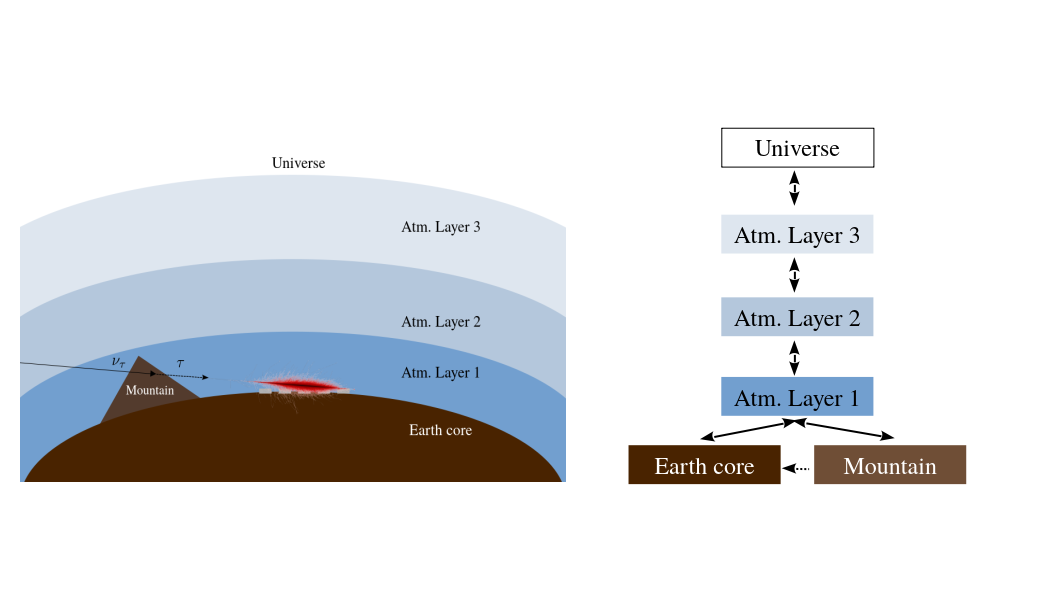}
\caption{On the left is shown an example of a possible experimental setup in CORSIKA 8. On the right, the tree structure corresponding to the adjacent environment. Figure adapted from \cite{Reininghaus:2019jxg, Dembinski:2020wrp}.}
\label{fig:env_example}
\end{figure}

One of the novelties present in CORSIKA 8 is the option of having complex geometries where multiple volumes with different density profiles and other properties, such as refractivity, can be combined. In particular, the ability to simulate showers in media with non-constant density is one thing that sets apart CORSIKA 8 from \textsc{Geant4}~\cite{agostinelli2003geant4}. An example of such flexibility is shown in Fig.~\ref{fig:env_example}. This level of customization is possible thanks to the volume tree CORSIKA 8 uses to represent the world. 

The user first defines the shape of each object present in the environment they want to create, currently spheres and cuboids. After the geometry of the volume is defined, the user will specify the physical properties of it. Lastly, once all the volumes have been defined and given properties, they need to be assembled in a tree. In this step it is where the user specifies the volume containment, and thus, how particles travel from one medium to another. In Fig.~\ref{fig:env_example} we make use of all the flexibility offered by CORSIKA 8, not only by having spheres fully contained in each other, but by adding a mountain (represented by a cuboid partially covered by the Earth core) which is contained in the first atmosphere layer and intersecting the Earth core.

\subsection{Comparison with CORSIKA 7 and \textsc{Geant4}}

Until now the simulations setups to run cross media showers in realistic scenarios were done by running the showers until they reached a new medium, saving the particles that reached that medium and starting a new simulation with them in a different setup \cite{de2022simulation, tueros2010tierras, motloch2016transition}. With a comparison of CORSIKA 8 against CORSIKA 7 plus \textsc{Geant4} we intend to verify the proper functioning of it and consolidate the results of the other frameworks.

For the comparison we will take advantage of the previously published results in \cite{de2022simulation}, where a 100 PeV proton shower was propagated through the atmosphere with CORSIKA 7 and intersected with an ice core at an altitude of 2.4 km, simulated by \textsc{Geant4}. The same layered atmosphere has been used in both programs (the MSIS-E-90 atmospheric model for South Pole on December 31, 1997) and the same ice density profile was used for the ice core. Due to the limitation of \textsc{Geant4} to only simulate homogeneous media, the density profile of the ice (eq.~\ref{eq:ice_density}) was implemented in multiple flat vertical layers of 1 cm thick of varying density. Taking advantage of the CORSIKA 8 capabilities we implement a continuous medium with spherical geometry. 

For CORSIKA 8 to propagate particles in a medium, the density function, $\rho(h)$, the conversion from length to grammage, $X(l)$, and the conversion from grammage to length, $l(X)$, need to be provided. The functions implemented for ice in this work were:
\begin{equation}
    \rho(h) = a_1+a_2\left[1-e^{-0.02(h^{\textrm{int}}-h)}\right]
\label{eq:ice_density}
\end{equation}
\begin{equation}
    X(l) = (a_1+a_2)l+\frac{a_2}{\cos(\theta)b} e^{-b(h^{\textrm{int}}-h_1)}(1-e^{b l \cos(\theta)})
\end{equation}
\begin{equation}
    l(X) = \frac{\frac{a_1+a_2}{c}W_0\left(b\cos\theta \exp{\left[(X/c-1)b\cos\theta\frac{c}{a_1+a_2}\right]}\frac{c}{a_1+a_2}\right)-(\frac{X}{c}-1)b\cos\theta}{\frac{a_1+a_2}{c}b\cos\theta}
\end{equation}
where $a_1 = 0.460 ~\mathrm{g\,cm}^{-3}$, $a_2 = 0.468 ~\mathrm{g\,cm}^{-3}$, $b=0.02 ~\mathrm{m}$, $h_1$ is the starting altitude of a particle track in meters, $h^{\textrm{int}}$ is the air-ice boundary altitude, $c=\frac{a_2}{b\cos\theta}e^{-b(h^{int}-h_1)}$ and $W_0(x)$ is the Lambert's W function. Lastly, for the environment, we also set the same uniform magnetic field, the magnetic field at the South Pole $\textbf{B}=(9.07, ~0, ~61.80)~\mu \textrm{T}$.

\begin{figure}[ht]
\centering
\subfigure{\includegraphics[width=0.49\linewidth]{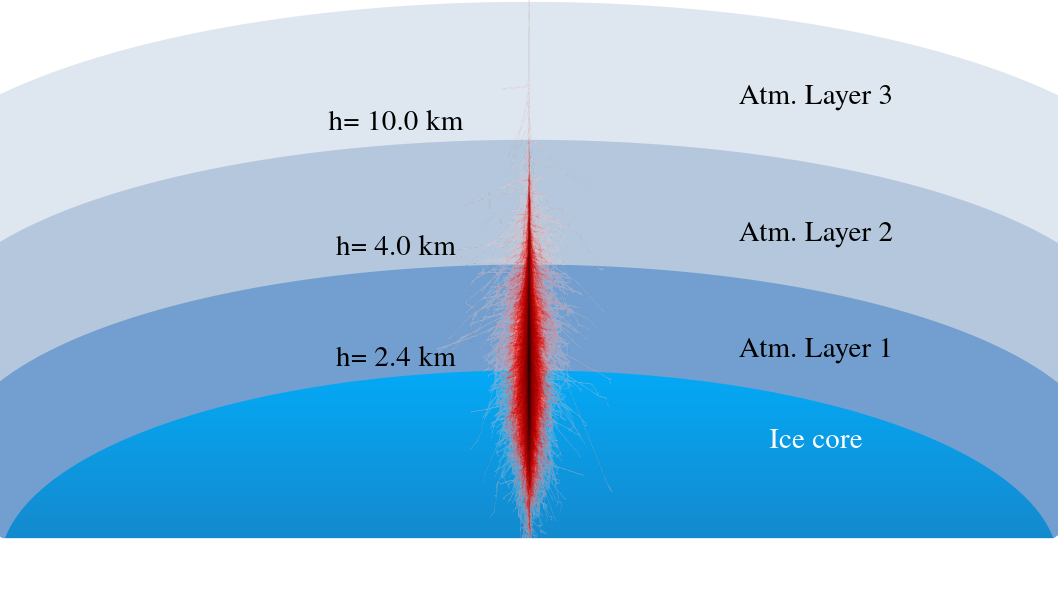}}
\subfigure{\includegraphics[width=0.49\linewidth]{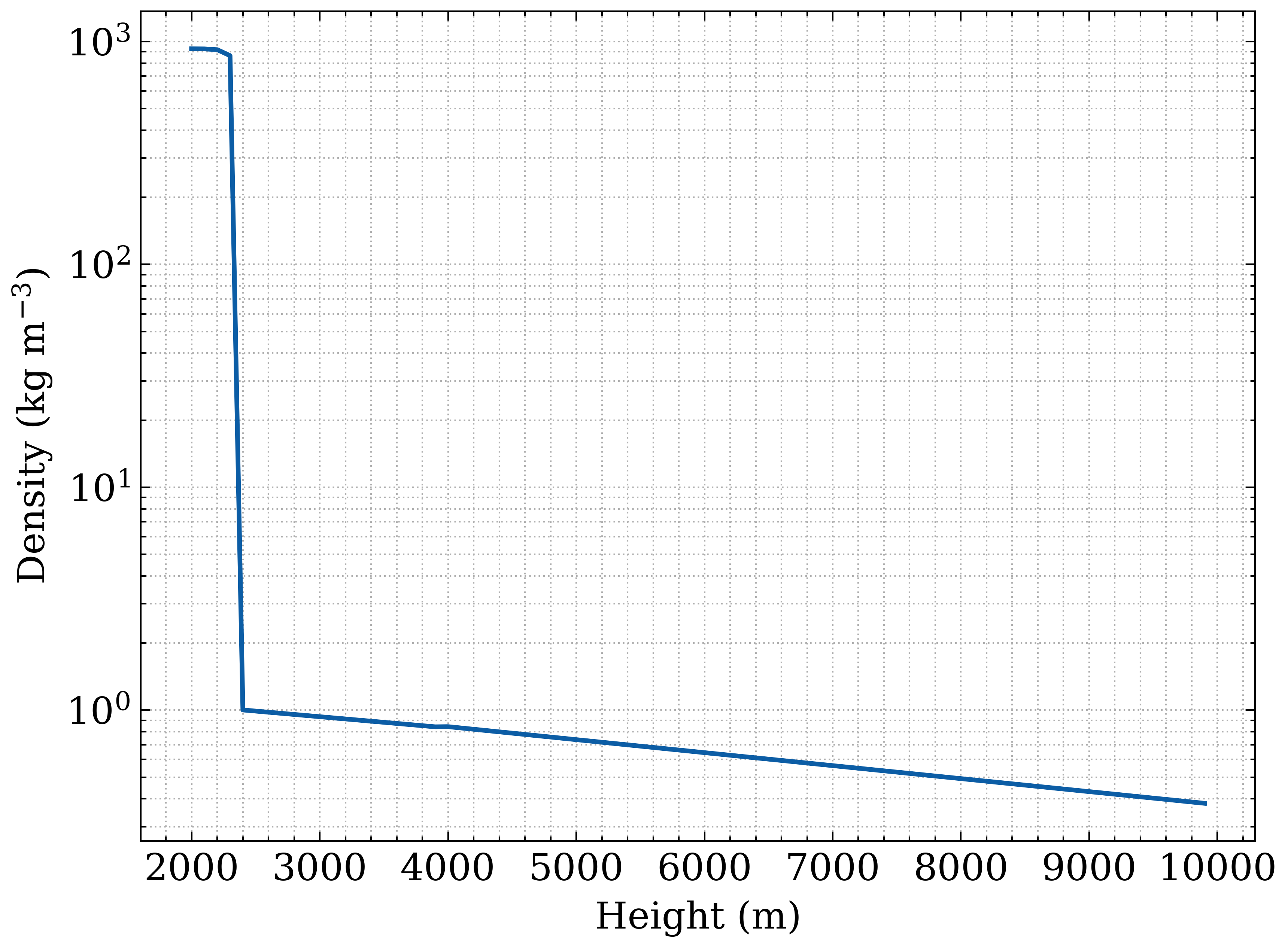}}
\caption{On the left a sketch of the geometry used for the comparison can be found. On the right, the vertical density profile is shown.}
\label{fig:cross_media_shower_geometry}
\end{figure}

For the interaction models there are some differences between both setups. Table \ref{table:interaction_models} summarizes all the different modules used for the interactions. The absence of hadronic interaction models in-ice for the \textsc{Geant4} setup chosen for the comparison means that in those showers there will be no re-hadronization in the ice. The re-hadronization is caused by the drastic increase in density (see Fig.~\ref{fig:cross_media_shower_geometry} right) when the shower enters the ice, which changes the relation between the interaction and decay length, favoring the hadronic interaction over the decay.

\begin{table}[ht]
\centering
\resizebox{\textwidth}{!}{
\begin{tabular}{cc|c|c|c|c|c|}
\cline{3-7}
\multicolumn{2}{c|}{}                                                         & Hadrons HE  & Hadrons LE                                              & Hadronic Decay                                                                         & Photohadronic                                              & EM                                                             \\ \hline
\multicolumn{1}{|c|}{C7} & Air                                                & QGSJETII-04 & \begin{tabular}[c]{@{}c@{}}GHEISHA\\ 2002d\end{tabular} & C7                                                                                     & QGSJETII-04                                                & EGS4                                                           \\ \hline
\multicolumn{1}{|c|}{G4} & Ice                                                & -           & -                                                       & \begin{tabular}[c]{@{}c@{}}G4DecayPhysics,\\ G4Radioactive\\ DecayPhysics\end{tabular} & -                                                          & \begin{tabular}[c]{@{}c@{}}G4EmStandard\\ Physics\end{tabular} \\ \hline
\multicolumn{1}{|c|}{C8} & \begin{tabular}[c]{@{}c@{}}Air,\\ Ice\end{tabular} & SIBYLL 2.3d & \begin{tabular}[c]{@{}c@{}}FLUKA\\ 2021.2\end{tabular}  & \begin{tabular}[c]{@{}c@{}}PYTHIA 8.245,\\ SYBILL 2.3d\end{tabular}                    & \begin{tabular}[c]{@{}c@{}}SOPHIA,\\ PROPOSAL\end{tabular} & PROPOSAL                                                       \\ \hline
\end{tabular}}
\caption{Interaction models used for the comparison between CORSIKA 7 (C7) plus \textsc{Geant4} (G4) and Corsika 8 (C8) for a cross-media shower between air and ice.}
\label{table:interaction_models}
\end{table}

Regarding the energy cuts, for CORSIKA 7, electrons, positrons, photons and pions had a kinetic energy threshold of 3 MeV, while hadrons and muons had it set to 300 MeV. For the \textsc{Geant4} ice shower no energy cuts are available, instead it uses a minimum interaction length, which was set to 1 mm (for the longitudinal profile, only particles above 3 MeV are counted). For the CORSIKA 8 simulation the cut for electrons, positrons and photons was set to 3 MeV, the muon cut to 300 MeV and the hadrons to 300 MeV (currently there is no option to have a specific cut for pions). For the the energy deposit plot the electron, positron and photon cut was lowered to 0.5 MeV to better adjust to the 1 mm interaction length cut in \textsc{Geant4}.

\begin{figure}[ht]
\centering
\subfigure[CORSIKA 8]{\includegraphics[width=0.49\linewidth]{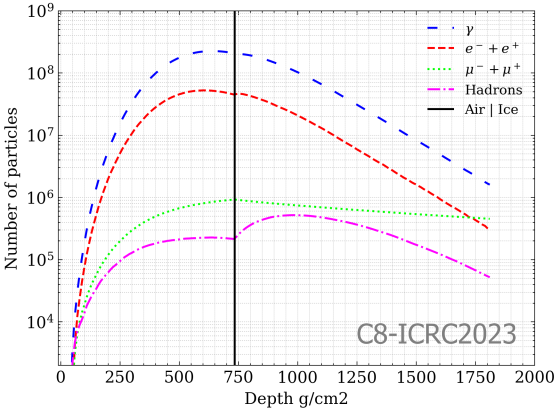}}
\subfigure[CORSIKA 7 \& \textsc{Geant4}]{\includegraphics[width=0.49\linewidth]{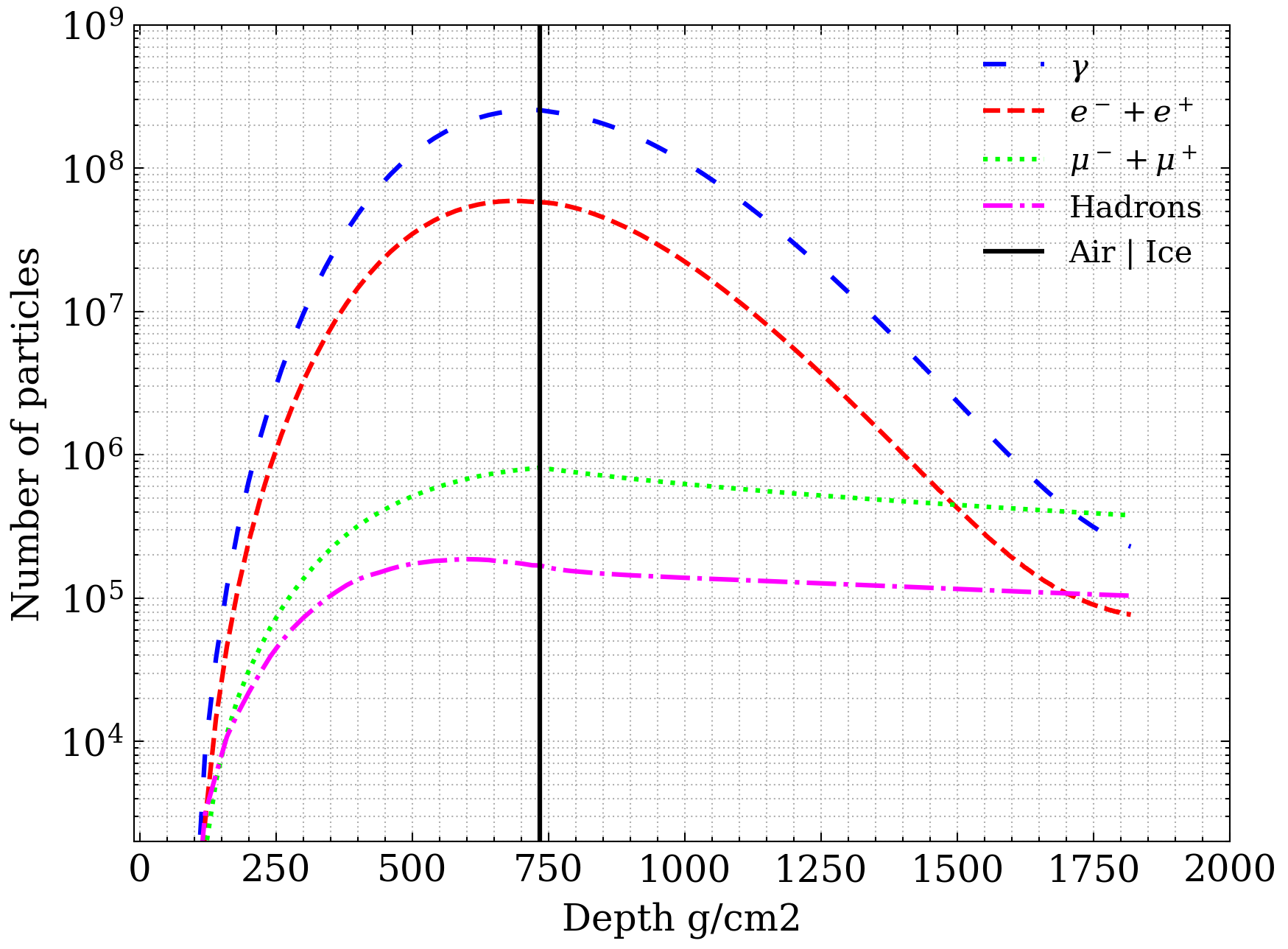}}
\caption{Longitudinal profile of a vertical 100 PeV proton cross-media shower that starts at the top of the atmosphere and intersects an ice layer located at an altitude of 2.4 km over sea level.}
\label{fig:comparison_longitudinal}
\end{figure}

In Fig.~\ref{fig:comparison_longitudinal} the longitudinal profile of a 100 PeV proton vertical air shower intersecting the Antarctic ice can be found for both setups, on the left CORSIKA 8 and on the right CORSIKA 7 plus \textsc{Geant4}. The overall agreement of the two simulation frameworks is good since we are only comparing one shower and different interaction models. As mentioned before, due to the lack of hadronic interactions in-ice for the chosen \textsc{Geant4} configuration, one cannot see the rehadronization in Fig.~\ref{fig:comparison_longitudinal} (right). CORSIKA 8 shows an increase in the number of hadrons when the showers intersects the ice due to the dominance of the hadronic interactions (with a partial contribution from the photohadronic interaction also) over the hadronic decays. The geometrical length  changes compared to the depth traversed by the shower. This rehadronization is also the cause for the small bump seen in electrons and photons after the interface, as $\pi^0$ are generated and decay to photons. The rehadronization effect has also been observed in other works like \cite{tueros2010tierras,ulrich2021hadron}.

\begin{figure}[ht]
\centering
\subfigure[CORSIKA 8]{\includegraphics[width=0.49\linewidth, trim={3cm 0cm 2.5cm 0cm},clip]{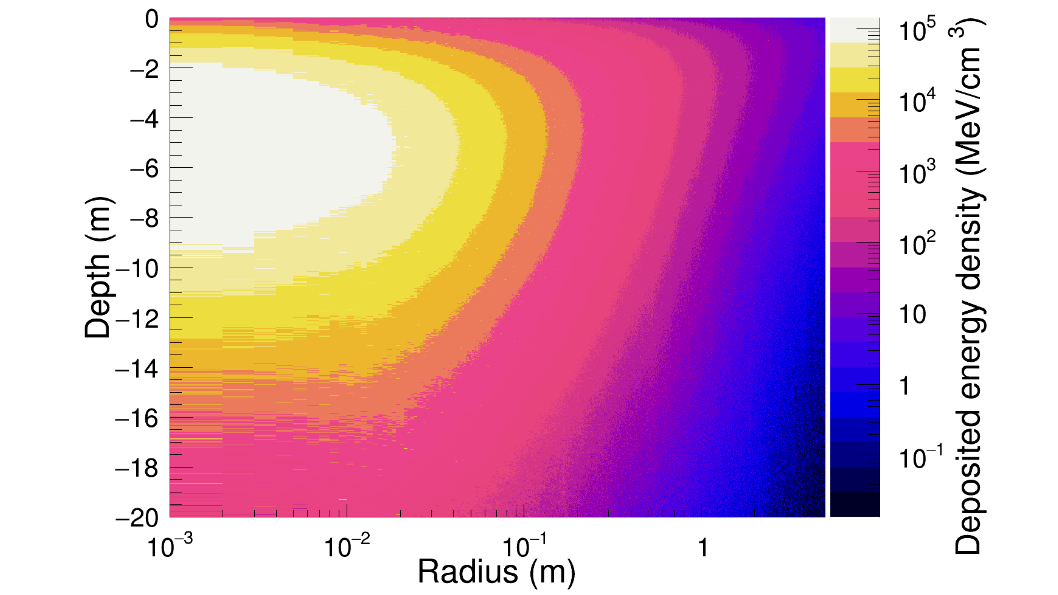}}
\subfigure[CORSIKA 7 \& \textsc{Geant4}]{\includegraphics[width=0.49\linewidth]{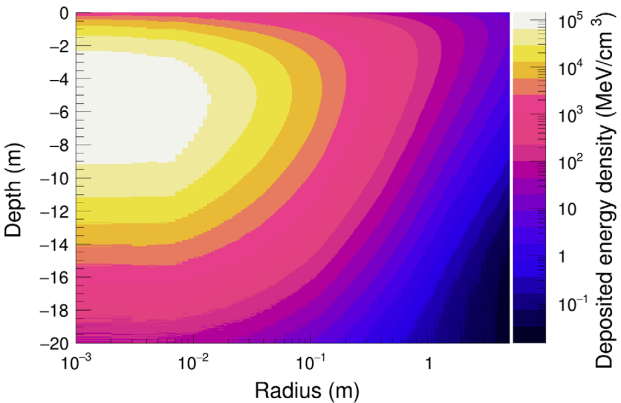}}
\caption{Radial energy deposit of a vertical 100 PeV proton cross-media shower that starts at the top of the atmosphere and intersects an ice layer located at an altitude of 2.4 km over sea level.}
\label{fig:comparison_energy_deposit}
\end{figure}

To check the physics of the cross-media shower we also calculate the radial energy deposit in-ice, which will test the longitudinal and radial development of the particles that arrive to the second medium. The results can be seen in Fig.~\ref{fig:comparison_energy_deposit}. The qualitative agreement is good between the two frameworks. A precise agreement between the two plots is difficult since the ice part of the shower is highly dependent on the evolution of the air shower (different amounts of energy going into the ice), which will never match one to one. The granularity seen in the CORSIKA 8 plot can be attributed to a higher and more aggressive thinning.

\section{Medium-specific propagators}
Having verified the correct development of cross-media showers in CORSIKA 8, an interesting application we can do is the simulation of the radio emission from the in-ice part of the shower. This is of particular importance for background rejection in experiments searching for neutrinos interacting in the ice through their radio emission \cite{Rice-Smith2022, Barwick:2022vqt}. We will do this in a simplified case to test the radio interface and its handling of multiple signal trajectories \cite{karastathis2023radio}. 

The media used for this example are the previous atmosphere plus an homogeneous ice core that starts at 2.4 km of altitude of density 0.46~$\mathrm{g\,cm}^{-3}$ with refractive index of 1.43. This way there are two straight optical paths the radio emission could follow: direct and reflected. Currently the radio interface does not accommodate phase shifts that can occur at reflections at the surface. For now, we only take the attenuation into account.

\begin{figure}[ht]
\centering
\includegraphics[width=0.6\linewidth]{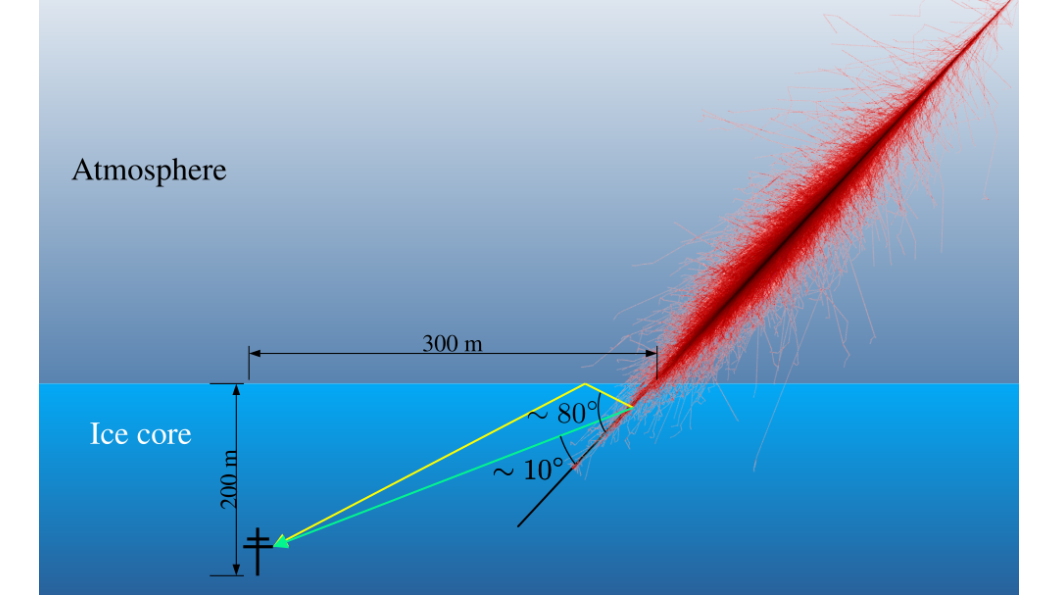}
\caption{Sketch of the geometry used for the simulation of the radio pulses in Fig.~\ref{fig:ice_pulse}. The antenna is placed at $(-300,~0,~-200)$ m with respect to the point of intersection between the shower axis and the ice core. Not to scale.}
\label{fig:radio_sketch}
\end{figure}

\begin{figure}[ht]
\centering
\subfigure[X polarization]{\includegraphics[width=0.49\linewidth]{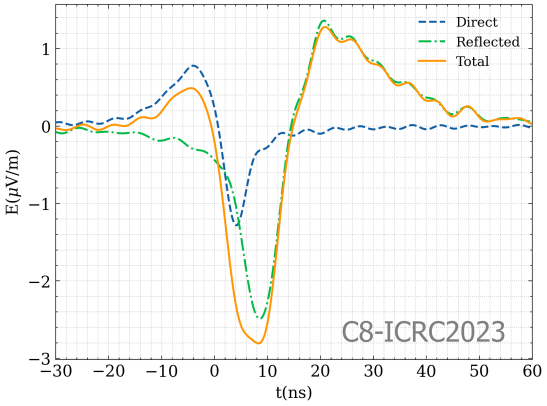}}
\subfigure[Z polarization]{\includegraphics[width=0.49\linewidth]{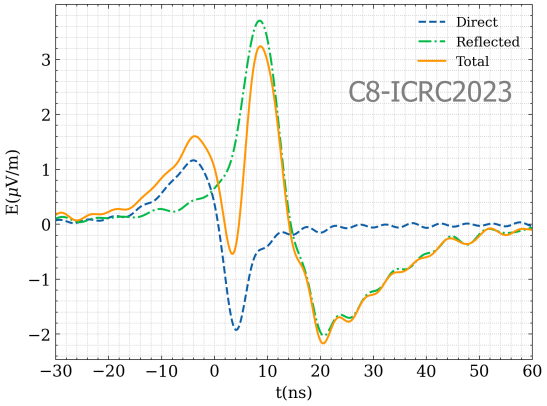}}
\caption{Radio pulse from the in-ice part of a cross-media shower that starts in the atmosphere and intersects an ice-core at 2.4 km of altitude. The shower was initiated by an 100 PeV proton at 45$^\circ$. A hard bandpass filter from 1 MHz to 2 GHz was applied to the pulse.}
\label{fig:ice_pulse}
\end{figure}

The geometry chosen is that of a 45$^\circ$ zenith angle 100 PeV proton shower starting at the top of the atmosphere and intersecting an ice-core Fig.~\ref{fig:radio_sketch}. The observer has been placed at $(-300,~0,~-200)$ m with respect to the point of intersection between the shower axis and the ice core. The radio pulse obtained is shown in Fig. \ref{fig:ice_pulse}. Since the observer is placed in the $\boldsymbol{xz}$ plane, no relevant emission is produced in the Y polarization\footnote{Due to the high density of ice, though having a mostly vertical magnetic field, no geomagnetic emission is produced.} and therefore is not shown. The X and Z polarizations both show pulses that are tens of nanoseconds wide. This is caused by the emission angles of both contributions \ref{fig:radio_sketch}. The Cherenkov angle for this ice is 45.63$^\circ$, so neither of the contributions is close to it, resulting in a less coherent pulse. It is also of interest to mention the small time displacement, of around 15 nanoseconds, between the direct and the reflected signal. The cause for it is that the shower development in ice, due to its high density, only lasts a few meters, making a small difference in the length of both signal paths.

\section{Summary}

The CORSIKA 8 framework for cross-media showers has been shown to be working correctly. In particular we have tested the specific scenario of an air shower that intersects the Antarctic ice and compared it with the results of CORSIKA 7 and \textsc{Geant4}. For this, an ice medium following the Antarctic Taylor Dome ice cap was implemented, together with a new module that calculates the radial energy deposit.

A good agreement between both simulation setups was found for a vertical 100 PeV proton shower as can be seen in Figs.~\ref{fig:comparison_longitudinal},\ref{fig:comparison_energy_deposit}. The main difference found between both codes being that of the re-hadronization in CORSIKA 8 due to the inclusion of hadronic interactions in-ice.

The correct functioning of cross-media showers in CORSIKA 8 allows to test new scenarios with an easier setup. As a first demonstration of what can be done, a simple example of only in-ice radio signal propagation, was implemented, showing both the possibility of selecting the medium where the propagator is calculated and the option of having multiple signal paths (direct and reflected). In particular, for radio, it is foreseen to continue this line of work and provide curved, cross media propagators that will also include propagation effects.

\section{Acknowledgements}
J. Ammerman-Yebra would like to acknowledge the funding received from the Spanish Research State Agency, Ministerio de Ciencia e Innovaci\'on/Agencia Estatal de Investigaci\'on PRE2020-092276. This research was funded by the Deutsche Forschungsgemeinschaft (DFG, German Research Foundation) – Projektnummer 445154105. This work has been supported by the the European Research Council under the EU-ropean Unions Horizon 2020 research and innovation programme (grant agreement No 805486).

\let\oldbibliography\thebibliography
\renewcommand{\thebibliography}[1]{%
  \oldbibliography{#1}%
  \setlength{\itemsep}{1pt}%
}

{\footnotesize
\bibliography{main}
}

\clearpage

\section*{The CORSIKA 8 Collaboration}
\small

\begin{sloppypar}\noindent
J.M.~Alameddine$^{1}$,
J.~Albrecht$^{1}$,
J.~Alvarez-Mu\~niz$^{2}$,
J.~Ammerman-Yebra$^{2}$,
L.~Arrabito$^{3}$,
J.~Augscheller$^{4}$,
A.A.~Alves Jr.$^{4}$,
D.~Baack$^{1}$,
K.~Bernl\"ohr$^{5}$,
M.~Bleicher$^{6}$,
A.~Coleman$^{7}$,
H.~Dembinski$^{1}$,
D.~Els\"asser$^{1}$,
R.~Engel$^{4}$,
A.~Ferrari$^{4}$,
C.~Gaudu$^{8}$,
C.~Glaser$^{7}$,
D.~Heck$^{4}$,
F.~Hu$^{9}$,
T.~Huege$^{4,10}$,
K.H.~Kampert$^{8}$,
N.~Karastathis$^{4}$,
U.A.~Latif$^{11}$,
H.~Mei$^{12}$,
L.~Nellen$^{13}$,
T.~Pierog$^{4}$,
R.~Prechelt$^{14}$,
M.~Reininghaus$^{15}$,
W.~Rhode$^{1}$,
F.~Riehn$^{16,2}$,
M.~Sackel$^{1}$,
P.~Sala$^{17}$,
P.~Sampathkumar$^{4}$,
A.~Sandrock$^{8}$,
J.~Soedingrekso$^{1}$,
R.~Ulrich$^{4}$,
D.~Xu$^{12}$,
E.~Zas$^{2}$

\end{sloppypar}

\begin{center}
\rule{0.1\columnwidth}{0.5pt}
\raisebox{-0.4ex}{\scriptsize$\bullet$}
\rule{0.1\columnwidth}{0.5pt}
\end{center}

\vspace{-1ex}
\footnotesize
\begin{description}[labelsep=0.2em,align=right,labelwidth=0.7em,labelindent=0em,leftmargin=2em,noitemsep]
\item[$^{1}$] Technische Universit\"at Dortmund (TU), Department of Physics, Dortmund, Germany
\item[$^{2}$] Universidade de Santiago de Compostela, Instituto Galego de F\'\i{}sica de Altas Enerx\'\i{}as (IGFAE), Santiago de Compostela, Spain
\item[$^{3}$] Laboratoire Univers et Particules de Montpellier, Universit\'e de Montpellier, Montpellier, France
\item[$^{4}$] Karlsruhe Institute of Technology (KIT), Institute for Astroparticle Physics (IAP), Karlsruhe, Germany
\item[$^{5}$] Max Planck Institute for Nuclear Physics (MPIK), Heidelberg, Germany
\item[$^{6}$] Goethe-Universit\"at Frankfurt am Main, Institut f\"ur Theoretische Physik, Frankfurt am Main, Germany
\item[$^{7}$] Uppsala University, Department of Physics and Astronomy, Uppsala, Sweden
\item[$^{8}$] Bergische Universit\"at Wuppertal, Department of Physics, Wuppertal, Germany
\item[$^{9}$] Peking University (PKU), School of Physics, Beijing, China
\item[$^{10}$] Vrije Universiteit Brussel, Astrophysical Institute, Brussels, Belgium
\item[$^{11}$] Vrije Universiteit Brussel, Dienst ELEM, Inter-University Institute for High Energies (IIHE), Brussels, Belgium
\item[$^{12}$] Tsung-Dao Lee Institute (TDLI), Shanghai Jiao Tong University, Shanghai, China
\item[$^{13}$] Universidad Nacional Aut\'onoma de M\'exico (UNAM), Instituto de Ciencias Nucleares, M\'exico, D.F., M\'exico
\item[$^{14}$] University of Hawai'i at Manoa, Department of Physics and Astronomy, Honolulu, USA
\item[$^{15}$] Karlsruhe Institute of Technology (KIT), Institute of Experimental Particle Physics (ETP), Karlsruhe, Germany
\item[$^{16}$] Laborat\'orio de Instrumenta\c{c}\~ao e F\'\i{}sica Experimental de Part\'\i{}culas (LIP), Lisboa, Portugal
\item[$^{17}$] Fluka collaboration
\end{description}

\end{document}